\documentclass[11pt,twocolumn,aip,apl,superscriptaddress,reprint]{revtex4-1}
\usepackage{graphicx,amsmath,amssymb,color,hyperref,epstopdf}

\begin{document}
\title{Optimal asymmetry of transistor-based terahertz detectors}

\author{Aleksandr Shabanov}
\affiliation{Moscow Institute of Physics and Technology, Dolgoprudny 141700, Russia}

\author{Maxim Moskotin}
\affiliation{Moscow Institute of Physics and Technology, Dolgoprudny 141700, Russia}
 
\author{Vsevolod Belosevich}
\affiliation{Physics Department, Moscow State Pedagogical University, Moscow 119435, Russia}
\affiliation{National Research University Higher School of Economics, Moscow 101000, Russia}

\author{Yakov Matyushkin}
\affiliation{Moscow Institute of Physics and Technology, Dolgoprudny 141700, Russia}
\affiliation{National Research University Higher School of Economics, Moscow 101000, Russia}

\author{Maxim Rybin}
\affiliation{Prokhorov General Physics Institute of the Russian Academy of Sciences, Moscow 119991, Russia}

\author{Georgy Fedorov}
\affiliation{Moscow Institute of Physics and Technology, Dolgoprudny 141700, Russia}

\author{Dmitry Svintsov}
\affiliation{Moscow Institute of Physics and Technology, Dolgoprudny 141700, Russia}

\begin{abstract}
Detectors of terahertz radiation based on field-effect transistors (FETs) are among most promising candidates for low-noise passive signal rectification both in imaging systems and wireless communications. However, it was not realised so far that geometric asymmetry of common FET with respect to source-drain interchange is a strong objective to photovoltage harvesting. Here, we break the traditional scheme and reveal the optimally-asymmetric FET structure providing the maximization of THz responsivity. We fabricate a series of graphene transistors with variable top gate position with respect to mid-channel, and compare their sub-THz responsivities in a wide range of carrier densities. We show that responsivity is maximized for input gate electrode shifted toward the source contact. Theoretical simulations show that for large channel resistance, exceeding the gate impedance, such recipe for responsivity maximisation is universal, and holds for both resistive self-mixing and photo-thermoelectric detection pathways. In the limiting case of small channel resistance, the thermoelectric and self-mixing voltages react differently upon changing the asymmetry, which may serve to disentangle the origin of nonlinearities in novel materials.

\end{abstract}

\maketitle
Sensitive detection of sub-terahertz and terahertz (THz) radiation is vital for security applications, defect inspection, radio-astronomy, and medical imaging~\cite{Dhillon_2017}. Currently, the main application niche for sub-THz detectors is wireless communications, where increase in carrier frequency promises a proportional increase in data transfer rates~\cite{Nagatsuma2016}. Recently, antenna-coupled field-effect transistors (FETs) have emerged as prospective candidates for passive sub-THz and THz detection~\cite{Knap2009,Bauer2016,Bauer2019}. Compared to active systems based on high-frequency amplifiers and followed by rectifiers, passive FET-based systems feature low power consumption and can operate above the cutoff frequency~\cite{Bauer2014c}. Compared to passive systems based on diodes~\cite{Liu2017c}, FETs are compatible with planar CMOS technology and do not require complex three-dimensional designs. The presence of extra control electrode in FETs simplifies the phase coherent (homodyne~\cite{Shur_APL_Homodyne} and heterodyne~\cite{Glaab_APL_Heterodyne}) detection. These advantages stimulate a booming research on THz FET detectors, including those based on novel materials~\cite{Vicarelli,Vitiello_ADMA_BlackPDetectors,Knap_NL_TI-detectors,Gayduchenko2018,Viti2019}, alternative mechanisms of current manipulation~\cite{Auton2017e,Gayduchenko2021}, and exploiting plasmonic effects~\cite{Knap2004,Muravev2012a,Bandurin2018}.

It has been scarcely realised that most FETs operating as THz detectors have an intrinsic deficiency lying in symmetry of the structure with respect to the interchange of source and drain. Once the signal is fed between source and drain, the photovoltage may build up only at finite bias~\cite{Xia2009,Gan2013}, which results in large Flicker noise. Finite zero-bias response for source-drain signal coupling appears also if a lateral $p-n$ junction is introduced in the channel which, however, requires extra ``doping gates''~\cite{Castilla2019}. Coupling the THz signal between source and gate, known as Dyakonov-Shur scheme~\cite{Dyakonov1996}, introduces an {\it asymmetry of signal feeding}, and thus results in the desired zero-bias photovoltage between source and drain~\cite{Knap2004a}. 

\begin{figure}[ht]
    \centering
    \includegraphics[width=0.9\linewidth]{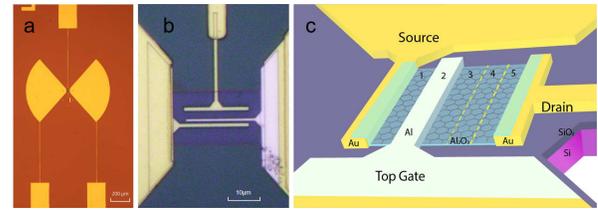}
    \caption{\label{fig:Scheme}Antenna-coupled graphene FET as terahertz detector (a) Optical image of the antenna-coupled device. Scale bar is 200~$\mu$m (b) Zoomed-in photograph of the gated channel. Scale bar is 10 $\mu$m. (c) Schematic of the FET-based detector with asymmetric gate location. Numbers indicate five gate positions used in our study.
    }
\end{figure}

To date, most research on THz FETs blindly copied the proposal of Dyakonov and Shur with geometrically symmetric channel~\cite{Knap2004a,Bandurin_dual,Zak_Roskos,Vicarelli,Vitiello_ADMA_BlackPDetectors,Lisauskas_2009_RationalDesign}. It was not attempted to enhance the responsivity by introducing extra {\it geometric asymmetry} that is achieved, most simply, by displacement of antenna-coupled gate electrode away from mid-channel. The necessity for geometric asymmetry was recently realised for THz emission applications~\cite{ElFatimy2010,Petrov2019}, but its role in detection process remain largely unexplored.

In this paper, we find an optimal geometric asymmetry of a FET-based antenna-coupled detector of THz radiation. We fabricate a series of graphene-based transistors with variable position of signal gate (Fig.~\ref{fig:Scheme}). Measurements of sub-THz photoresponse at room and liquid nitrogen temperatures show $\sim 5$-fold responsivity enhancement of a structure with proximized source and gate, compared to that with proximized drain and gate. 

 \begin{figure*}[ht]
    \centering
    \includegraphics[width=0.9\linewidth]{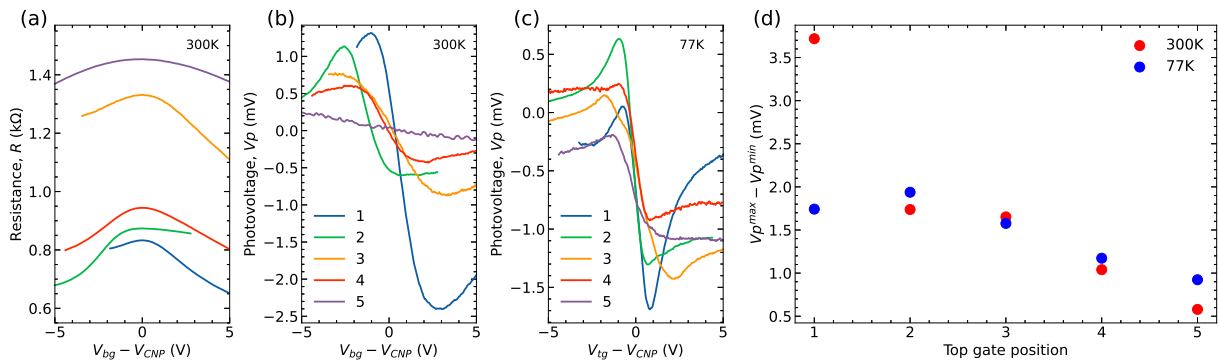}
    \caption{
    (a) Dependence of structures' channel resistance on bottom gate voltage for devices with different gate position. All curves are shifted such that charge neutrality voltages $V_{\rm CNP}$ coincide. 
    (b) Dependence of photovoltage on bottom gate voltage, obtained for the same devices upon illumination with sub-THz radiation ($f=130$ GHz). 
    (c) Dependence of photovoltage on top gate voltage for the same devices at temperature of liquid nitrogen.
    (d) Photovoltage swing upon variation of charge carrier density vs gate position obtained at $T=77$ K and $T=300$ K
    }
    \label{fig:Exp_swing}
\end{figure*}

We substantiate the obtained results theoretically, assuming two common rectification mechanisms: photo-thermoelectric (Seebeck) effect at metal-graphene junctions~\cite{Song2011,Cai2014} and resistive self-mixing in the gated channel~\cite{Sakowicz2011,Vicarelli}. Once the graphene channel resistance $R_{\rm ch}$ exceeds the ac gate impedance $|Z_C|$, both rectification channels benefit from source-shifted asymmetry, in accordance with experimental data. Our result has a simple interpretation if one assumes thermoelectric origin of photoresponse: co-location of source and gate increases the power density released at source metal-graphene junction, and enhances the average thermal gradient across the structure. The trend for resistive self-mixing does not admit that simple explanation, and actually changes for small $R_{\rm ch}$. In the limiting case $|Z_C| \gg R_{\rm ch}$, the self-mixing signal is reduced with shifting the gate toward the source,  while Seebeck signal stays roughly constant. Thus, experiments with variable gate location can shed light on dominant THz rectification pathways in graphene~\cite{Bandurin_dual,Zak_Roskos} and emerging 2D materials~\cite{Vitiello_ADMA_BlackPDetectors}.

The key challenge toward the quantitative comparison of photoresponse in CVD-graphene FETs lies in fluctuations of characteristics from one device to another. To circumvent this, we have fabricated nominally identical devices from a single millimetre-scale layer of graphene. It was grown in a home-made cold-wall reactor on copper foil~\cite{Rybin2016,Rybin_Synthesis}. Grapene was wet-transferred onto an oxidized Si substrate with SiO$_2$ dielectric of thickness $d_{b} = 500$~nm~\cite{Li2009}. SLG FET channel was patterned to a rectangular shape with length $L = 4$~$\mu$m and width $W = 20$~$\mu$m by e-beam lithography (PMMA mask) and etching in O$_2$ plasma and then supplied with two small metal contacts. These source and drain electrodes were made of gold (40~nm) with vanadium adhesion layer (5~nm). To maintain stability of graphene channel with respect to atmospheric contaminants, an aluminum oxide cover was deposited using electron-beam evaporation with a thickness of $d_t = 100$~nm and lift-off technique, which would also serve as a top gate dielectric. The small source electrode was coupled by a sleeve of a dipole antenna, and the other sleeve was deposited over the top gate [Fig.~\ref{fig:Scheme} (a,b)]. The top gate for various FETs was placed in five sequential positions, from that proximized to the source (labelled as {\#1}) to that in immediate vicinity to the drain (labeled as {\#5}) [Fig.~\ref{fig:Scheme} (c)]. For each device, we have performed electrical DC characterization; the silicon substrate covered with 500 nm of silica was used as a back gate. The dependences of the sample resistance on the gate voltage are presented in Fig.~\ref{fig:Exp_swing}(a); for all devices there is a maximum resistivity corresponding to the charge neutrality point (CNP) of graphene.

The main experiment consists in measurements of photovoltage $V_{\rm p}$ generated in FETs upon their illumination with sub-THz radiation of frequency $f = 130$ GHz. In this experiment, the detector on substrate was mounted on a holder with silicon lens arranged in a cryostat. Radiation was fed from backward wave oscillator with  power $P\sim 1$ mW calibrated with Golay cell.

The results of photovoltage measurements for various gate locations are summarized in Fig.~\ref{fig:Exp_swing}. The dependence of $V_{\rm p}$ on back gate voltage has a well-recognized anti-symmetric shape. Such shape is typical for both photo-thermoelectric and resistive self-mixing rectification pathways. A remarkable property of the recorded characteristics lies in their swing. Indeed, the photovoltage response for rightmost top gate position is faint and noisy, while the characteristic responsivity $R_{V} = V_{\rm p}/P$ hardly reaches $0.5$ V/W. The swing gradually goes up upon shifting the gate to the source, and reaches its maximum value for top gate proximized to the source. The characteristic responsivity for such geometry already reaches $2.5$ V/W, which is five times larger than for the rightmost gate position. 

\begin{figure*}[ht]
	\centering
	\includegraphics[width=0.9\linewidth]{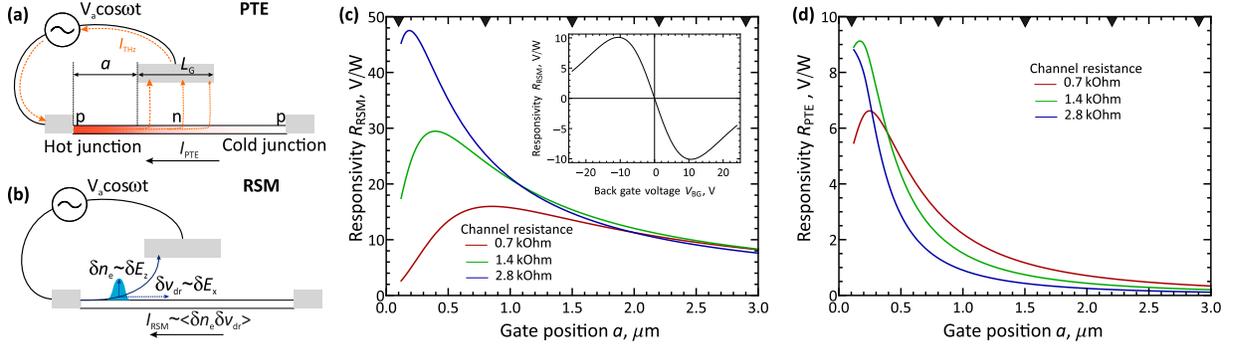}
	\caption{\label{fig:RSMnPTE}(a,b) Schematic representation of photothermoelectric (a) and resistive self-mixing (b) mechanisms in FETs with antenna coupled between source and gate. In (a), asymmetric distribution of AC current $I_{\rm THz}$ leads to overheating of source side of the junction, and to the thermal diffusion current toward the source. In (b), the transverse component of AC electric field induces extra charge carriers $\delta n_e$ and enhances the channel conduction. These carriers are subsequently dragged by in-plane component of the field. The time-averaged current $I_{\rm RSM}$ is directed toward the source. (c,d) Calculated responsivities of resistive self-mixing (c) and photo-thermoelectric effects in devices with variable gate-to-source separation $a$ calculated for different channel resistances $R_{\rm ch}$. All curves calculated for fixed Fermi energy $E_F=100$ meV and variable momentum relaxation times $\tau$ linked to resistivity $\rho$ via Drude formula $\rho = m(E_F) / n_e e^2 \tau$. Inset in (c) shows the dependence of responsivity on back gate voltage for device with $a=0.5$ $\mu$m.} 
\end{figure*}

To provide a simple figure-of-merit for devices with various asymmetry, we compare their photovoltage swings $\Delta V_{\rm p}$ being the differences between maximum and minimum photovoltages recorded upon variation of carrier density, Fig.~\ref{fig:Exp_swing}(d). Such a figure is insensitive to the unavoidable shifts of CNP due to large-scale doping fluctuations within CVD graphene layer. The maximum and minimum are reached at two sides of CNP, corresponding to weak hole/electron doping of the channel. Again, the trend toward increased photoresponse with shifting the gate toward the source is recognised in photovoltage swing data.

The trend toward increased photoresponse for source-shifted feeding gate persists upon cooling the sample to liquid nitrogen temperature, as shown in Fig. \ref{fig:Exp_swing} (c). The data for sample with gate position \# 1 fall out of this trend, probably due to sample-to-sample variations. Instructively, the characteristic responsivity at $T=77$~K is almost the same as at $T=300$~K. Weak sensitivity of electrical properties to the temperature in CVD graphene may be explained by a large number of intrinsic defects~\cite{Adam2007} obtained during growth and wet transfer.

We now turn to theoretical interpretation of the data assuming the two most common rectification mechanisms in graphene channel, the photo-thermoelectric effect (PTE) and the resistive self-mixing (RSM), illustrated in Fig.~\ref{fig:RSMnPTE} (a), (b). The PTE voltage emerges due to asymmetric heating of the device by ac electric current circulating between source and top gate~\cite{Cai2014,Bandurin_dual}. The metal-graphene junction at the source is strongly heated, while the junction at the drain remains at lower electron temperature. The difference in thermoelectric voltages generated by hot and cold junctions results in overall finite photovoltage between source and drain. The RSM emerges upon induction of charge carriers in the channel by transverse component of ac electric field, and subsequent drag of induced carriers by the longitudinal component~\cite{Sakowicz2011}.

Our quantitative model of photo-response is based on finding ac electric potential distribution $\varphi_\omega(x)$ in the channel induced by antenna voltage $V_a \cos\omega t = (e^{-i\omega t}+ e^{i\omega t}) V_a/2 $. This voltage can be bound to incoming THz power $P$ via antenna radiation resistance $Z_{\rm rad}$, $P = V_a^2/8Z_{\rm rad}$~\cite{Sanchez1978}. In the gated section, the potential is governed by the telegrapher's equation
\begin{equation}
\label{Telegrapher}
    \varphi_\omega + k^2\frac{\partial^2\varphi_\omega}{\partial x^2}=\frac{V_a}{2},
\end{equation}
where $k=(\sigma/i\omega C)^{1/2}$ is the wave vector of overdamped 2d plasmons~\cite{Dyakonov1996}, $\sigma$ is the conductivity of graphene sheet, and $C$ is the top gate-to-channel capacitance per unit area. The left ungated section of length $a$ is modelled as a lumped resistor, $\varphi_\omega(0) = a \varphi_\omega'(0)$. The rightmost ungated section does not affect the photoresponse as almost no ac current is flowing into the drain, thus $\varphi'_\omega(L_g) = 0$. The geometric asymmetry introduced by shift of the top gate is now reflected in the asymmetry of the boundary conditions for electric oscillation problem. 

The distribution of electric field $E_\omega(x) = -\partial \varphi_\omega/\partial x$ obtained from Eq.~(\ref{Telegrapher}) is subsequently used to calculate the RSM and PTE photoresponse. The respective voltages $V_{\rm RSM}$ and $V_{\rm PTE}$ are given by~\cite{Bandurin2018,Bandurin_dual}
\begin{equation}
\label{URSM}
    V_{\rm RSM}=2 \frac{d \sigma (V_{\rm tg})}{d V_{\rm tg}}\frac{1}{C\omega} {\rm Im}\int_0^{L_G}{ E_\omega(x) \frac{\partial E_\omega^*(x)}{\partial x}dx},
\end{equation}
\begin{equation}
\label{UPTE}
    V_{\rm PTE}=[S_{\rm ch}(V_{\rm bg}) - S_{\rm cont}][T_d -T_s].
\end{equation}
Above, $S_{\rm ch}(V_{\rm bg})$ and $S_{\rm cont}$ are the Seebeck coefficients in the back-gated channel and graphene in immediate contact with metal. For gold-contacted CVD graphene, the contacts are generally $p$-doped ($S_{\rm cont} >0$), which results in overall shift of all responsivity curves to negative photovoltages. $T_s$ and $T_d$ in Eq.~(\ref{UPTE}) are the electron temperatures at the metal-induced p-n junctions in graphene located near the source and the drain, these are found by solving heat conduction equation with ac Joule heating as a source:
\begin{equation}\label{Th_eqn}
\frac{\partial}{\partial x}\left( \chi_e \frac{\partial T}{\partial x} \right) - c_e\frac{T-T_0 }{\tau_{\varepsilon}}=2 \sigma |E_\omega|^2,
\end{equation}
above $c_e$ is the electronic heat capacitance, $\chi_e$ is the electronic thermal conductivity, and $\tau_\varepsilon$ is the energy relaxation time associated with supercollisions and emission of substrate phonons. We have assumed $\tau_{\varepsilon} = 1$ ps~\cite{Low2012}, the trends in PTE with gate shifts do not depend on specific value. Equation (\ref{Th_eqn}) was solved assuming fixed temperature $T_0$ (base cryostat temperature) at source and drain. The temperatures of interest $T_s$ and $T_d$ are evaluated at small distance $\delta L = 125$ nm from source an drain. This distance is the typical length of metal-induced $p-n$ junction acting as a thermoelectric generator~\cite{Cai2014}.

With the developed model, we have calculated the gate voltage dependences of photoresponse and found good qualitative agreement with experiment [inset in Fig.~\ref{fig:RSMnPTE}(c)]. Further on, we fixed numerically the value of carrier density in the channel and inspected the changes in photoresponse with varying the top gate position at various momentum relaxation times [Fig.~\ref{fig:RSMnPTE}(c), (d)]. Again, in good agreement with experimental data, the calculated photoresponse is growing as the position of the top gate is shifted toward the source for both rectification pathways. Remarkably, RSM and PTE photovoltages react differently on changes in the relaxation time. While RSM is weakly sensitive to $\tau_p$, the PTE signal benefits from increasing the momentum relaxation time.

\begin{figure}
    \centering
    \includegraphics[width=0.9\linewidth]{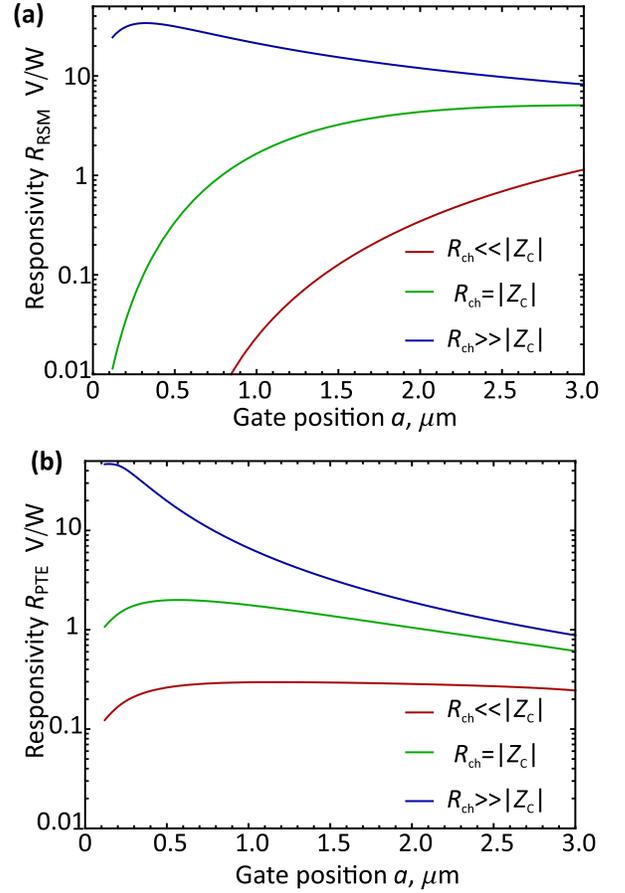}
    \caption{\label{fig:limits} Calculated dependences of resistive self-mixing (a) and photo-thermoelectric (b) responsivities on gate-to-source separation $a$ for various ratios of channel resistance $R_{\rm ch}$ and top gate capacitor impedance $Z_C = (i\omega C_{\rm tg})^{-1}$. While for $R_{\rm ch} \gg |Z_C|$ both mechanisms are maximised at small $a$, the trend is reversed for $R_{\rm ch} \ll |Z_C|$. Blue and green curves plotted for $Z_C = 100$ $\Omega$, red curve -- for $Z_C = 1$ k$\Omega$. The value of $R_{\rm ch}$ is 100 $\Omega$ for red and green curves, and 1.4 k$\Omega$ for blue curve.
    }
\end{figure}

In complement to the developed theory, it is possible to give a simple and intuitive interpretation of the measured data. The main dimensionless parameter governing the detector operation is the ratio of gate-to-channel capacitor impedance and gated channel resistance $p = [\omega C W L_G R_G]^{-1} \equiv k^2 L_G^2$. Our detector operates in high-resistance mode wherein $p \ll 1$ (indeed, $R_G \sim 1$ k$\Omega$ and $(\omega C L_G W)^{-1}\sim 100$ $\Omega$ in our device). Therefore, the THz voltage generated by the antenna drops mainly on graphene channel but not on the gate-channel capacitance. The power dissipated on ungated section $P \approx V_a^2/2R_u$ is growing as its length $a$ is decreased. The power density (per unit length) is increased even faster, $q \propto a^{-2}$ which results in an abrupt increase in junction heating and, hence, PTE voltage with shifting the gate to the source. The RSM effect should also benefit from left-ward shifting of the gate. Whence $kL \ll 1$, it is possible to estimate the longitudinal electric field under the gate as $V_a / a$. Reducing $a$ leads to stronger potential variations in the gated section and stronger self-mixing effects.

It is instructive that in the other limiting case of large gate-channel capacitance, $R_{\rm ch} \ll |Z_C|$, the behaviour of RSM and PTE effects with shifting the signal gate is very different (Fig.~\ref{fig:limits}). In such a limit, the amplitude of current through graphene channel $I_{\rm THz}$ is fixed by large gate impedance. The power dissipated in the ungated section $P_u \approx 2 |I_{\rm THz}|^2 R_u$ is proportional to the ungated length $a$, while the power density setting the temperature is independent of $a$. The numerically calculated variations of PTE response in this limit with shifting the gate are slight and appear due to non-local character of heat transport equation. Contrary to the PTE, the RSM signal goes down with reducing $a$. 

In principle, the experiment with variable gate position in the limit of large capacitive impedance may help to conclude whether the dominant THz detection mechanism in graphene is RSM or PTE~\cite{Bandurin_dual}. Such problem can be hardly resolved with simple gate-dependtent measurements. Indeed, the dependences of both mechanisms on carrier density are indistinguishable and follow the derivative of conductivity with respect to Fermi energy. The experiment with variable gate position is also applicable to other 2d materials and heterostructure FETs, where the origins of photoresponse are debated~\cite{Vitiello_ADMA_BlackPDetectors,Viti2019}.

Our interpretation of data was based on specific boundary conditions for ac current and voltage at the FET terminals. These imply fixed voltage at the antenna-coupled terminals, and zero current at the drain. The latter condition is approximate, as it neglects the leakage current through the fringing capacitance between the gate and the drain $C_{gd}$~\cite{Lisauskas_2009_RationalDesign}. At the same time, this capacitance is reduced once we reduce the gate-drain separation $a$, and so does the leakage current. This may be another reason contributing to enhanced photoresponse of FETs with close source and gate.

To conclude, we have shown that the optimal asymmetric structure of antenna-coupled terahertz detecting FET is the one with source proximized to the feeding gate. Considering the FET as two rectifying contacts located at the source and drain junctions, we may say that placement of the gate near the source concentrates the electromagnetic energy at one of the rectifiers, leaving the other intact. Following such arguments, we may suggest that 'source-shifted asymmetry' of FET-based detectors would be beneficial also for rectification by hydrodynamic non-linearities of electron fluid~\cite{Dyakonov1996,Principi_PseudoEuler}, non-linearities of Schottky/$p-n$ junctions of the contacts~\cite{Ryzhii2006c}, and photocurrent generation by photovoltaic effect~\cite{Echtermeyer2011}. 

%The photovoltage response demonstrated by our graphene FETs was induced by radiation of frequency $f = 130$~GHz generated by a backward wave oscillator with an output power of 1~mW. Radiation power was measured with a Golay cell. For radiation detection we used a NI DAQ system with a high resolution analog-to-digital converter. The device was placed onto the flat surface of a truncated ball silicon lens fixed inside a home-made holder. The connection between device contacts and holder pads was carried out with thin (25~$\mu$m) aluminum wire bondings by an ultrasonic wire bonder. The holder with the graphene FET experimental device was placed inside a cryostat with a THz-transparent HDPE window. Radiation entered the cryostat through the transparent window, then passed throughout the lens to focus onto the surface of our device, where it was coupled with the antenna. The diameter of the radiation beam ($\sim$ 3~mm) is larger than the size of the device, therefore, the effects associated with uneven illumination of the sample can be neglected.

{\bf{Supplementary material.}} See supplementary material for resistance and photovoltage data in a broad range of gate voltages.

{\bf{Acknowledgement.}} Experimental work of MM, VB, and GF was supported by grant \# 21-72-20050 of the Russian Science Foundation. YM, GF and VB acknowledge support of the RFBR grant \# 21-52-12041 (graphene post-synthesis processing and device design). Theoretical work of AS and DS was supported by grant \# 21-79-20225 of the Russian Science Foundation.

{\bf{Data availability statement.}} The data that support the findings of this study are available from the corresponding author upon reasonable request.

\bibliography{References} %Prints bibliography

\end{document}